\documentstyle[12pt]{article}
\begin{document}
\hoffset = -1truecm
\voffset = -2truecm
\topmargin = -1.2 cm

\title{\bf
Quasiparticle Many-Body Dynamics of Highly Correlated Electronic Systems
}
\begin{centering}
\author{
{\bf A. L. Kuzemsky \thanks{E-mail: kuzemsky@thsun1.jinr.dubna.su}}\\
\normalsize Bogolubov Theoretical Laboratory, \\
\normalsize Joint Institute for Nuclear Research, \\
\normalsize 141980 Dubna (Moscow Region), Russia}
\date{\empty}
\end{centering}
\maketitle
\begin{abstract}
The self-consistent theory of the correlation effects in Highly Correlated
Systems(HCS)  is presented. The novel Irreducible Green's Functions(IGF)
method
is discused in detail for the Hubbard model and random Hubbard model.
The interpolative
 solution for the quasiparticle spectrum, which is valid for both the
atomic and band limit is obtained.
The (IGF) method permits to calculate the quasiparticle spectra of
many-particle systems with the complicated spectra and strong interaction
in a very natural and compact way. The essence of the method deeply related
 with the notion of the Generalized Mean Fields (GMF), which determine
the elastic scattering corrections. The inelastic scattering corrections
leads to the damping of the quasiparticles and are the main topic of the
present consideration. The calculation of the damping has been done in a
self-consistent way for  both limits. For the random Hubbard model the
weak coupling case has been considered and the self-energy operator has
been calculated using the combination of the IGF method and Coherent
Potential Approximation (CPA). The other applications of the method to
s-f model, Anderson model, Heisenberg antiferromagnet, electron-phonon
interaction models and quasiparticle
tunneling are discussed briefly.
\end{abstract}
\newpage
\section{Introduction}
The study of the Highly Correlated Electron Systems has attracted
much attention recently, especially after discovery of copper oxide
superconductors and the new class of heavy fermion
compounds, coexisting with magnetism~\cite{kuz1} - \cite{dag6}.
Although much work has been performed during last years it is worthy
to remind that
the investigation of the excitations in many-body systems has been one of the most
important and interesting subject for last few decades. The quantum field
theoretical techniques have been widely applied to statistical treatment of
a large number of interacting particles. Many-body calculations are often
done for model systems of statistical mechanics using perturbation
expansion.
The basic procedure in many-body theory~\cite{mah7} is to find relevant
unperturbed
Hamiltonian and then take into account the small perturbation operator. This
procedure, which work well for the weakly interacting systems, needs the
suitable reformulation for the many-body systems with complicated spectra
and strong interaction. For many practically interesting cases the
standard schemes of perturbation  expansion must be reformulated
greatly~\cite{hur8} - \cite{tar12}.
The most characteristic feature of the recent advancement in basic research
on electronic properties of solids is development of variety of the new
class
of materials with unusual properties: high $T_c$ superconductors, heavy
fermion
compounds, diluted magnetic semiconductors etc. Contrary to the simple
metals,
where the fundamentals very well known and the electrons can be represented
in a way such that they weakly interact with each other(c.f.~\cite{theo13}),
in these materials
the electrons interact strongly and moreover their spectra are complicated,
i.e. have many branches etc. This gives rise to interesting phenomena such
as magnetism, metal-insulator transition in oxides, heavy fermions etc.,
but the understanding of what is going on is in many cases only partial if
exist at all. Therefore the theoretical studies of the Highly Correlated
Electron Systems (HCS) are very important and actual.
A principle
importance of of these studies is concerned with a fundamental problem
of electronic solid state theory, namely with the tendency of 3d electrons
in TMC and 4f electrons in rare-earth metal compounds (REC) and alloys to
exhibit both localized and delocalized behaviour. The interesting electronic
and magnetic properties of these substances are intimately related to this
dual behaviour of electrons. In spite of experimental and theoretical
achievements~\cite{kuz1} - \cite{dag6}, still it remains much to
be understood concerning
such systems.
Recent theoretical investigations of HCS have brought forth significant
variety of the approaches which are trying to solve these controversial
problems. It seems appropriate to point out that a number of
perturbation-theory or mean-field theory approaches which have been
proposed in the past few years, are in fact questionable or inadequate.
In order to match such a trend we need to develop a systematic theory of
the Highly Correlated Systems, to describe from the first principles of
the condensed matter theory and statistical mechanics the physical
properties of this class of the materials. In the present paper we will
present the approach which allows one to describe completely
the quasi-particle spectra with damping in a very natural way.
This approach has
been suggested as essential for various many-body systems and we believe
that it bear the real physics of Highly Correlated Systems~\cite{kuz14},
\cite{kuz15}.
The essence of our
consideration of the dynamical properties of many-body system with strong
interaction is related closely with the field theoretical approach and
use the advantage of the Green's functions language and the Dyson
equation. It is possible to say that our method tend to emphasize the
fundamental and central role of the Dyson equation for the single-particle
dynamics of the many-body systems at finite temperature.\\
Just this point differ our IGF approach from the complimentary
many-body  approach which is based on the moment expansion for the
spectral functions. It was developed in a very detail by  W.Nolting
~\cite{nol16a} - ~\cite{nol20a}.
\section{Irreducible Green's Functions Method}
In this Section, we will discuss briefly the novel nonperturbative
approach for description of the many-body dynamics of the HCES. At this
point it is worthwhile to underline that it is essential to apply an
adequate
method in order to solve a concrete physical problem; the final solution
should contain a correct physical reasoning in a most natural way. The
list of many-body techniques that have been applied to strongly correlated
systems is extensive. The problem of adequate description of many-body
dynamics for the case of very strong Coulomb correlations has been
explicitly
raised by Anderson, who put the direct question: ``... whether a real
many-body
theory would give answers radically different from the Hartree-Fock
results?"~\cite{and16} (c.f.~\cite{hald17}). The formulation of the Anderson
model\cite{and16}
and closely related Hubbard model~\cite{hub18}, \cite{kan19} dates really a
better
understanding of the electronic correlations in solids, especially if the
relevant electrons are modelled better by tight-binding
approximation~\cite{lieb20},\cite{lieb21}.
Both of the models, Anderson and Hubbard, are often referred to as simplest
models of magnetic metals and alloys. This naive perception contradicts the
enormous amount of papers which has been publishing during the last decades
and devoted to attacking the Anderson/Hubbard model by many refined
theoretical techniques. As is well known now, the simplicity of the
Anderson/Hubbard model manifest itself in the dynamics of a two-particle
scattering. Nevertheless, as to the true many-body dynamics, there is
still no simple and  compact  description.  In  this  paper  it  will  be
attempted
to justify the use of a novel Irreducible Green's Functions (IGF) for the
interpolation solution of the single-band Hubbard model and other basic
solid state models as s-f model~\cite{mar22}, \cite{mar23}, Anderson
model~\cite{kuz24},\cite{kuz25}, Heisenberg antiferromagnet~\cite{kuz26}
and  strong electron-phonon interaction model in modified tight-binding
approximatiom(MTBA) for normal and superconducting
metals~\cite{kuz27} and alloys~\cite{wys28}, \cite{wys29}.
A number of other
approaches has been proposed and the our approach is in many respect an
additional and incorporate the logic of development of the many-body
techniques.
The
considerable progress in studying the spectra of elementary excitations
and thermodynamic properties of many-body systems has been for most part
due to the development of the temperature dependent Green's Functions
methods. We have developed the helpful reformulation of the two-time GFs
method which is especially adjusted~\cite{kuz30} for the correlated fermion
systems on a lattice. The very important concept of the whole method are
the {\bf Generalized Mean Fields}. These GMFs have a complicated structure
for the strongly correlated case and are not reduced to the functional of
the mean densities of the electrons, when we calculate excitations
spectra at finite temperature.
To clarify the foregoing, let us consider the retarded GF of the form
\begin{equation}
G^{r} = <<A(t), B(t')>> = -i\theta(t - t')<[A(t)B(t')]_{\eta}>, \eta = \pm
1.
\end{equation}
As an introduction of the concept of IGFs let us describe the main ideas
of this approach in a symbolic form. To calculate the retarded GF
$G(t - t')$ let us write down the equation of motion for it:
\begin{equation}
\omega G(\omega) = <[A, A^{+}]_{\eta}> + <<[A, H]_{-}\mid A^{+}>>_{\omega}.
\end{equation}
The essence of the method is as follows~\cite{kuz14}. It is based on the
notion of the {\it ``IRREDUCIBLE"} parts of GFs (or the irreducible parts of
the operators, out of which the GF is constructed) in term of which it is
possible, without recourse to a truncation of the hierarchy of equations
for the GFs, to write down the exact Dyson equation and to obtain an
exact analytical representation for the self-energy operator. By definition
we introduce the irreducible part {\bf (ir)} of the GF
\begin{equation}
^{ir}<<[A, H]_{-}\vert A^{+}>> = <<[A, H]_{-} - zA\vert A^{+}>>.
\end{equation}
The unknown constant z is defined by the condition (or constraint)
\begin{equation}
<[[A, H]^{ir}_{-}, A^{+}]_{\eta}> = 0
\end{equation}
From the condition (4) one can find:
\begin{equation}
z = \frac{<[[A, H]_{-}, A^{+}]_{\eta}>}{<[A, A^{+}]_{\eta}>} =
 \frac{M_{1}}{M_{0}}
\end{equation}
Here $M_{0}$ and $M_{1}$ are the zeroth and first order moments of the
spectral density. Therefore, irreducible GF (3) are defined so that it
cannot be reduced to the lower-order ones by any kind of decoupling. It
is worthy to note that the irreducible correlation functions are well
known in statistical mechanics. In the diagrammatic approach the
irreducible vertices are defined as the graphs that do not contain
inner parts connected by the $G^{0}$-line. With the aid of the definition
(3) these concepts are translating into the language of retarded and
advanced GFs. This procedure extract all relevant (for the problem under
consideration) mean field contributions and puts them into the generalized
mean-field GF, which here are defined as
\begin{equation}
G^{0}(\omega) = \frac{<[A, A^{+}]_{\eta}>}{(\omega - z)}.
\end{equation}
To calculate the IGF $ ^{ir}<<[A, H]_{-}(t), A^{+}(t')>>$ in (2), we have
to write the equation of motion after differentiation with respect to the
second time variable $t'$. The condition (4) remove the inhomogeneous term
from this equation and is the very crucial point of the whole approach. If
one introduces an irreducible part for the right-hand side operator as
discussed above for the ``left" operator, the equation of motion (2) can be
exactly rewritten in the following form
\begin{equation}
G = G^{0} + G^{0}PG^{0}.
\end{equation}
The scattering operator $P$ is given by
\begin{equation}
P = (M_{0})^{-1}\quad ^{ir}<<[A, H]_{-}\vert[A^{+}, H]_{-}>>^{ir}
(M_{0})^{-1}.
\end{equation}
The structure of the equation (7) enables us to determine the self-energy
operator $M$, in complete analogy with the diagram technique
\begin{equation}
P = M + MG^{0}P.
\end{equation}
From the definition (9) it follows that we can say that the self-energy
operator $M$ is defined as a proper (in diagrammatic language ``connected")
part of the scattering operator $M = (P)^{p}$. As a result, we obtain the
exact Dyson equation for the thermodynamic two-time Green's Functions:
\begin{equation}
G = G^{0} + G^{0}MG,
\end{equation}
which has well known formal solution of the form
$$G = [ (G^{0})^{-1} - M ]^{-1}.$$
Thus, by introducing irreducible parts of GF (or the irreducible parts of
the operators, out of which the GF is constructed) the equation of motion
(2) for the GF can be exactly (but using constraint (4)) transformed into
Dyson equation for the two-time thermal GF. This is very remarkable result,
which deserve the underlining, because of the traditional form of the GF
method did not included namely this point.
The projection operator technique~\cite{for31}
has essentially the same philosophy, but with using the constraint (4) in
our approach we emphasize the fundamental and central role of the Dyson
equation for the calculation of the single-particle properties of the
many-body systems. It is important to note, that for the retarded and
advanced GFs the notion of the proper part is symbolic in
nature~\cite{kuz14}.
However, because of the identical form of the equations for the GFs for all
three types (advanced, retarded and causal), we can convert in each stage
of calculations to causal GFs and, thereby, confirm the substantiated nature
of definition (9)! We therefore should speak of an analog of the Dyson
equation. Hereafter we will drop this stipulation, since it will not cause
any misunderstanding. It should be emphasized that scheme presented above
give just an general idea of the IGF method. The specific method of
introducing
IGFs depends on the form of operator $A$, the type of the Hamiltonian and
the
conditions of the problem. The general philosophy of the IGF method lies in
the separation and identification of elastic scattering effects and
inelastic ones. This last point is quite often underestimated and both
effects are mixed. However, as far as the right definition of quasiparticle
damping is concerned, the separation of elastic and inelastic scattering
processes is believed to be crucially important for the many-body systems
with complicated spectrum and strong interaction. Recent paper~\cite{rie32}
emphasizes especially that the anomalous damping of electrons (or holes)
distinguishes cuprate superconductors from ordinary metals.
From a technical point
of view the elastic (GMF) renormalizations can exhibit a quite non-trivial
structure. To obtain this structure correctly, one must construct the full
GF from the complete algebra of the relevant operators and develop a
special projection procedure for higher-order GF in accordance with a
given algebra. The Hubbard model is a very suitable tool for the applying
of this approach~\cite{kuz30},\cite{kuz33}.
\section{ Hubbard Model}
The model Hamiltonian which is usually referred to as Hubbard Hamiltonian
\begin{equation}
H = \sum_{ij\sigma}t_{ij}a^{+}_{i\sigma}a_{j\sigma} +
U/2\sum_{i\sigma}n_{i\sigma}n_{i-\sigma}
\end{equation}
includes the intraatomic Coulomb repulsion $U$ and the one-electron hopping
energy $t_{ij}$. The electron correlation forces electrons to localize in
the atomic orbitals, which are modelled here by the complete and orthogonal
set of the Wannier wave functions $[\phi({\vec r} -{\vec R_{j}})]$. On the
other hand, the kinetic energy is reduced when electrons are delocalized.
The
main difficulty of the right solution of the Hubbard model is the necessity
to taking into account of the both these effects simultaneously. Thus, the
Hamiltonian (11) is specified by two parameter: $U$ and effective electron
bandwidth
$$\Delta = (N^{-1}\sum_{ij}\vert t_{ij}\vert^{2})^{1/2}.$$
The band energy of Bloch electrons $\epsilon(\vec k)$ is defined as follows
$$t_{ij} = N^{-1}\sum_{\vec k}\epsilon(\vec k)
\exp[i{\vec k}({\vec R_{i}} -{\vec R_{j}}],$$
where the $N$ is the number of the lattice sites. It is convenient to count
the energy from the center of gravity of the band, i.e.
 $t_{ii} = \sum_{k}\epsilon(k) = 0$. The effective electron bandwidth
 $\Delta$ and Coulomb intrasite integral $U$ define completely the
 different
 regimes in 3 dimension depending on parameter $\gamma = \Delta/U$. It is
 usually a rather difficult task to find interpolation solution for the
 dynamical properties of the Hubbard model. To solve this problem with a
 reasonably accuracy and describe correctly an interpolating solution from
 ``band" limit ($\gamma \gg 1$) to ``atomic" limit ($\gamma \rightarrow 0$)
 one need more sophisticated approach than usual procedures
 which have been developed for description of the interacting electron-gas-
 problem. We evidently have to to improve the early Hubbard's theory taking
 account of variety of possible regimes for the model depending on
electronic
 density, temperature and values of $\gamma$. The single-electron GF
 \begin{equation}
 G_{ij\sigma}(\omega) = <<a_{i\sigma} \vert a^{+}_{j\sigma}>> =
 N^{-1}\sum_{\vec k}G_{\sigma}({\vec k}, \omega)
 \exp[-i{\vec k}({\vec R_{i}} -{\vec R_{j}})],
 \end{equation}
 which has been calculated by Hubbard~\cite{hub18}, \cite{hub34}, has the
 characteristic two-pole functional structure
 \begin{equation}
 G_{\sigma}(k, \omega) = [ F_{\sigma}(\omega) - \epsilon(k)]^{-1}
 \end{equation}
 where
 \begin{equation}
 F^{-1}_{\sigma}(\omega) = \frac{\omega - (n^{+}_{-\sigma}E_{-} +
 n^{-}_{-\sigma}E_{+}) - \lambda}{(\omega - E_{+} - n^{-}_{-\sigma}\lambda)
 (\omega - E_{-} - n^{+}_{-\sigma}\lambda) - n^{+}_{-\sigma}n^{-}_{-\sigma}
 \lambda^{2}}
 \end{equation}
 and$ \lambda$ is the certain function which depends on parameters of the
 Hamiltonian. If $ \lambda$ is small ($\lambda \rightarrow 0$) then
expression
 (14) take the form:
 $$ F^{-1}_{\sigma}(\omega) \approx
 \frac{n^{-}_{-\sigma}}{\omega - E_{-} - n^{+}_{-\sigma}\lambda} +
 \frac{n^{+}_{-\sigma}}{\omega - E_{+} - n^{-}_{-\sigma}\lambda},$$
 which correspond to the two shifted subbands with the gap
 $$ \omega_{1} - \omega_{2} = (E_{+} - E_{-}) +
 (n^{-}_{-\sigma} - n^{+}_{-\sigma})\lambda =
 U + \lambda 2n^{+}_{-\sigma}. $$
 Here $n^{+} = n$ and $n^{-} = 1 - n$; $E_{+} = U$, $E_{-} = 0$. If
 $\lambda$ is very big then we obtain
 $$F^{-1}_{\sigma}(\omega) \approx \frac{\lambda}
 {[(\omega - E_{-})n^{-}_{-\sigma} + (\omega - E_{+})n^{+}_{-\sigma}]
 \lambda} = \frac{1}{\omega - ( n^{+}_{-\sigma}E_{+} -
 n^{-}_{-\sigma}E_{-})}.$$
 This latter solution correspond to the single band, centered at the
 energy $\omega \approx n^{+}_{-\sigma}U$. \\
 The two- pole functional structure of the single-particle GF is very
 easy to understand within formalism which describe the motion of
 electrons in binary alloys~\cite{hub34}, \cite{beeb35}. If one introduce
 the two types of the scattering potentials $t_{\pm} \approx
 (\omega - E_{\pm})^{-1}$ then the two kinds of the t-matrix $ T_{+}$ and
 $T_{-}$ appears which satisfy the following system of equations:
 $$T_{+} = t_{+} + t_{+}G^{0}_{++}T_{+} + t_{+}G^{0}_{+-}T_{-}$$\\
 $$T_{-} = t_{-} + t_{-}G^{0}_{--}T_{-} + t_{-}G^{0}_{-+}T_{+},$$
where $G^{0}_{\alpha\beta}$ is the bare propagator between the sites with
the energies $E_{\pm}$. The solution of this system has the following form
\begin{eqnarray}
T_{\pm} = \frac{t_{\pm} + t_{\pm}G^{0}_{\pm}t_{\pm}}
{(1 - t_{+}G^{0}_{++})(1 - t_{-}G^{0}_{--}) -
G^{0}_{-+}G^{0}_{+-}t_{+}t_{-}}
 =\nonumber\\
 \frac{t^{-1}_{\mp} + G^{0}_{\pm}}
{( t^{-1}_{+} - G^{0}_{++} )( t^{-1}_{-} - G^{0}_{--}) -
 G^{0}_{-+}G^{0}_{+-}}.
\end{eqnarray}
Thus, by comparing this functional two-pole structure and well-known
``Hubbard III" solution~\cite{hub34}\\
$$\Sigma_{\sigma}(\omega) = \omega - F_{\sigma}(\omega)$$
it is possible to identify the ``scattering corrections" and ``resonance
broadening corrections" in the following way:\\
$$F_{\sigma}(\omega) = \frac{\omega(\omega - U) -
(\omega - Un_{-\sigma})A_{\sigma}(\omega)}
{\omega - U(1 - n_{-\sigma}) - A_{\sigma}(\omega)}$$\\
$$A_{\sigma}(\omega) = Y_{\sigma}(\omega) + Y_{-\sigma}(\omega)
- Y^{*}_{-\sigma}(U - \omega)$$\\
$$Y_{\sigma} = F_{\sigma}(\omega) - G^{-1}_{0\sigma}(\omega);
G_{0\sigma}(\omega) = N^{-1}\sum_{k}G_{k\sigma}(\omega)$$\\
If we put $A_{\sigma}(\omega) = 0$ we immediately obtain the ``Hubbard I"
solution~\cite{hub18}. The ``alloy analogy" approximation correspond to
$A_{\sigma}(\omega) \approx Y_{\sigma}(\omega)$. Note, that the
``Hubbard III" self-energy operator $\Sigma_{\sigma}(\omega)$ is local,
i.e. do not depend on quasimomentum. The another drawback of this solution
is very inconvenient functional representation of the elastic and inelastic
scattering processes. The conceptually new approach to the theory of very
strong but finite electron correlation for Hubbard model has been proposed
by Roth~\cite{roth36}. She clarified microscopically the origination of the
two-pole solution of the single-particle GF, what was the very unusual fact
from the point of view of the standard Fermi-liquid approach, showing that
the naive one-electron approximation of the band structure calculations
is not valid for the description of the electron correlations in HCS . Thus
the use of sophisticated many-body technique is required for the calculation
of the excitation spectra at finite temperature. This last point should be
underlined, because of the suitable modification of the Density Functional
Approximation~\cite{gun37}, \cite{ter38},\cite{sva39} could
give the reasonable
description of the ground-state properties of HCS. We shall show here,
following the papers~\cite{kuz30},\cite{kuz33} that
the use of the IGF method permit to improve substantially both solutions,
Hubbard's and Roth's, by defining the correct Generalized Mean Fields for
the Hubbard model.
\section{Hubbard Model. Weak Correlation}
The concept of the GMFs and the relevant algebra of operators from which GFs
are constructed are the central ones to our treatment of electron
correlation
in solids. It will be convenient (and much more shorter) to discuss these
concepts for weakly and strongly correlated cases separately. For the first
time
we must to construct the suitable state vector space of the many-body
system~\cite{bog40}. The fundamental assumption implies that  the
states of a system of interacting particles can be expanded in terms of
the states of non-interacting particles~\cite{bog40}. This concept
originate in perturbation theory and finds support for weakly interacting
many-particle systems(c.f.~\cite{mar5}).
For the strongly correlated case this approach
needs the the suitable reformulation (cf. \cite{hal41}) and namely in this
point the right definition of the GMFs is vital.
Let us consider the weakly correlated Hubbard model (11). In many respect
this
case is similar to the ordinary interacting electron gas but with very
local, singular interaction. It will be shown below that the usual creation
$a^{+}_{i\sigma}$ and annihilation $a_{i\sigma}$ second quantized operators
with the properties
$$a^{+}_{i}\Psi^{(0)} = \Psi^{(1)}_{i} ; a_{i}\Psi^{(1)} = \Psi^{(0)}$$
$$a_{i}\Psi^{(0)} = 0 ; a_{j}\Psi^{(1)}_{i} = 0,    ( i \not= j)$$
are suitable variables for the description of the considering systems. Here
$\Psi^{(0)}$ and $\Psi^{(1)}$ are the vacuum and single-particle states
respectively.
The question now is how to describe our system in terms of the
quasiparticles.
 For a translationally invariant system, to describe the low-lying
excitations
of the system in terms of quasiparticles~\cite{bog40}, one has to choose
eigenstates such that they all correspond to definite momentum. For the
single-band Hubbard model (11) the exact transformation reads
$$a_{{\vec k}\sigma} = N^{-1/2}\sum_{i}\exp(-i{\vec k}{\vec R_{i}})
a_{i\sigma}$$
Note, that for degenerate bands model the more general transformation
is necessary. Then the Hubbard Hamiltonian (11) in the Bloch vector state
space are given by
\begin{equation}
H = \sum_{k\sigma}\epsilon(k)a^{+}_{k\sigma}a_{k\sigma} +
U/2N\sum_{pqr}\sum_{\sigma}a^{+}_{p+r-q\sigma}a_{p\sigma}a^{+}_{q-\sigma}a_{r-\sigma}
\end{equation}
If the interaction is weak, the algebra of the relevant operators is very
simple: it is an algebra of the non-interacting fermion system
($a_{k\sigma}, a^{+}_{k\sigma}, n_{k\sigma} = a^{+}_{k\sigma}a_{k\sigma}$).
For the calculation of the electronic quasiparticle spectrum of the
Hubbard model in this limit let us consider the single-electron GF, which
are defined as
\begin{eqnarray}
G_{k\sigma}(t - t') = <<a_{k\sigma}, a^{+}_{k\sigma}>> =
-i\theta(t - t')<[a_{k\sigma}(t), a^{+}_{k\sigma}(t')]_{+}> = \nonumber\\
1/2\pi \int_{-\infty}^{+\infty} d\omega\exp(-i\omega t)G_{k\sigma}(\omega)=
\nonumber\\
1/2\pi \int_{-\infty}^{+\infty}d\omega\exp(-i\omega t)
1/2\pi \int_{-\infty}^{+\infty}\frac{d\omega'}{\omega - \omega'}
(\exp(\beta \omega') + 1)A_{k\sigma}(\omega')
\end{eqnarray}
where $\beta = (kT)_{-1}$ and $A_{k\sigma}(\omega)$ is the spectral
intensity. The equation of motion for the Fourier transform of the GF
$G_{k\sigma}(\omega)$ has the form
\begin{equation}
(\omega - \epsilon_{k})G_{k\sigma}(\omega) = 1 +
U/N\sum_{pq}<<a_{k+p\sigma}a^{+}_{p+q-\sigma}a_{q-\sigma}
\vert a^{+}_{k\sigma}>>_{\omega}
\end{equation}
Let us introduce, by definition, an ``irreducible" GF in the following
way
\begin{eqnarray}
^{ir}<<a_{k+p\sigma}a_{p+q-\sigma}a_{p+q-\sigma}a_{q-\sigma}
\vert a^{+}_{k\sigma}>>_{\omega} =\nonumber\\
<<a_{k+p\sigma}a^{+}_{p+q-\sigma}a_{q-\sigma} \vert
a^{+}_{k\sigma}>>_{\omega}
-\delta_{p, 0}<n_{q-\sigma}>G_{k\sigma}
\end{eqnarray}
The irreducible (ir) GF in (19) is defined in such a way that it cannot
be reduced to GF of lower order with respect to the number of fermion
operators by an arbitrary pairing of operators or, in another words, by
any kind of decoupling. Substituting (19) in (18) we obtain
\begin{equation}
G_{k\sigma}(\omega) = G^{MF}_{k\sigma}(\omega) + G^{MF}_{k\sigma}(\omega)
U/N\sum_{pq}{}^{ir}<<a_{k+p\sigma}a^{+}_{p+q-\sigma}a_{q-\sigma}
\vert a^{+}_{k\sigma}>>_{\omega}
\end{equation}
Here we have introduced the notations
\begin{equation}
G^{MF}_{k\sigma}(\omega) = (\omega - \epsilon(k\sigma))^{-1} ;
\epsilon(k\sigma) = \epsilon(k) +U/N\sum_{q}<n_{q-\sigma}>
\end{equation}
In this paper, for brevity, we confine ourself by considering the
paramagnetic
solutions only, i.e. $<n_{\sigma}> = <n_{-\sigma}>$.
In order to calculate the higher-order GF on the r.h.s. of (20) we
have to write the equation of motion obtained by means of differentiation
with respect to the second variable $t'$. Constraint (4) allows us to remove
the inhomogeneous term in this equation for
$\frac{d}{dt'}^{ir}<<A(t), a^{+}_{k\sigma}(t')>>$. \\
For the Fourier components, this is written in the form
\begin{eqnarray}
(\omega - \epsilon(k)){}^{ir}<<A\vert a^{+}_{k\sigma}>>_{\omega} =
<^{ir}[A, a^{+}_{k\sigma}]_{+}> +\nonumber\\
U/N\sum_{rs}{}^{ir}<<A\vert a^{+}_{r-\sigma}a_{r+s-\sigma}a_{k+s\sigma}>>_
{\omega}.
\end{eqnarray}
The anticommutator in (22) is calculated on the basis of the definition
of the irreducible part
\begin{eqnarray}
<[^{ir}(a_{k+p\sigma}a^{+}_{p+q-\sigma}a_{q-\sigma}), a^{+}_{k\sigma}]_{+}>
=
\nonumber\\
<[a_{k+p\sigma}a^{+}_{p+q-\sigma}a_{q-\sigma} -
<a^{+}_{p+q-\sigma}a_{q-\sigma}>a_{k+p\sigma}, a^{+}_{k\sigma}]_{+}> = 0
\end{eqnarray}
If one introduces irreducible part for the r.h.s. operators by analogy
with expression (19), the equation of motion (20) can be exactly
rewritten in the form (7)
\begin{equation}
G_{k\sigma}(\omega) = G^{MF}_{k\sigma}(\omega) + G^{MF}_{k\sigma}(\omega)
P_{k\sigma}(\omega)G^{MF}_{k\sigma}(\omega)
\end{equation}
where we have introduced the following notation for the operator $P$ (8)
\begin{eqnarray}
P_{k\sigma}(\omega) = \frac{U^{2}}{N^{2}}\sum_{pqrs}D^{ir}_{k\sigma}
(p, q\vert r, s, ;\omega) =\nonumber\\
= \frac{U^{2}}{N^{2}}\sum_{pqrs}{}^{ir}<<a_{k+p\sigma}a^{+}_{p+q-\sigma}
a_{q-\sigma}\vert a^{+}_{r-\sigma}a_{r+s-\sigma}a^{+}_{k+s\sigma}>>^{ir}_
{\omega}
\end{eqnarray}
To define the self-energy operator according to the (9) one must separate
the ``prop\-er" part by the following way
\begin{eqnarray}
D^{ir}_{k\sigma}(p, q\vert r, s;\omega) =
 L^{ir}_{k\sigma}(p, q\vert r, s;\omega)
\nonumber\\
+ \frac{U^{2}}{N^{2}}\sum_{r's'p'q'}L^{ir}_{k\sigma}(p, q\vert r's';\omega)
G^{MF}_{K\sigma}(\omega)D^{ir}_{k\sigma}(p', q'\vert r, s;\omega)
\end{eqnarray}
Here $L^{ir}_{K\sigma}(p, q\vert r, s;\omega)$ is the ``proper" part of the
GF $D^{ir}_{k\sigma}(p, q \vert r, s;\omega)$, which in accordance with the
definition (19) cannot be reduced to the lower-order one by any type of
decoupling. Using (9) we find
\begin{equation}
G_{k\sigma} = G^{MF}_{k\sigma}(\omega) + G^{MF}_{k\sigma}(\omega)
M_{\sigma}(k, \omega)G_{k, \sigma}(\omega)
\end{equation}
Equation (27) is the Dyson equation for the single-particle two-time
thermal GF. According to (10) it has the formal solution
\begin{equation}
G_{k\sigma}(\omega) = [\omega -\epsilon(k\sigma) -M_{\sigma}(k,
\sigma)]^{-1}
\end{equation}
where the self-energy operator $M$ is given by
\begin{eqnarray}
M_{\sigma}(k, \omega) = \frac{U^{2}}{N^{2}} \sum_{pqrs}L^{ir}_{k\sigma}
(p, q \vert r, s;\omega) =\nonumber\\
 \frac{U^2}{N^2}{} \sum_{pqrs}{} ^{ir}<<a_{k+p\sigma}a^{+}_{p+q-\sigma}
a_{q-\sigma} \vert a^{+}_{k+s\sigma}a^{+}_{r-\sigma}a_{r+s-\sigma}>>^{ir}
\end{eqnarray}
The latter expression (29) is an exact representation (no decoupling
has been made till now) for the self-energy in terms of higher-order
GFs up to second order in $U$ (for the consideration of the higher order
equations of motion see Ref. ~\cite{kuz15}). Thus, in contrast to the
standard equation-of-motion approach the determination of the full GF has
been reduced to the calculation of the mean-field GF $G^{MF}$ and the
self-energy operator $M$. The main reason for this method of calculation
is that the decoupling is only introduced into self-energy operator, as
it will shown in a detail below. The formal solution of the Dyson equation
(28) define the right reference frame for the formation of the
quasiparticle spectrum due to the its own (formal solution)
correct functional structure. In the standard equation-of-motion approach
such a structure could be lost by using decoupling approximations
{\bf before} arriving to the correct functional structure of the formal
solution of the Dyson equation. This is a crucial point of the IGF method.
The energies of the electronic states in the mean-field approximation are
given by the poles of $G^{MF}$ (21). Now let us consider the damping
effects and finite lifetimes. To find an explicit expression for
self-energy $M$ (29), we have to evaluate approximately the higher-order
GF in (21). It will be shown below that the IGF method can be used to
derive the damping in a self-consistent way simply and more generally than
other formulations. First, it is convenient to write down the GF in (29)
in terms of correlation functions by using the well-known spectral
theorem~\cite{bog40}:
\begin{eqnarray}
<<a_{k+p\sigma}a^{+}_{p+q-\sigma}a_{q-\sigma} \vert
a^{+}_{k+s\sigma}a^{+}_{r-\sigma}a_{r+s-\sigma}>>_{\omega} =\nonumber\\
{1 \over 2\pi}\int_{-\infty}^{+\infty}{d\omega' \over \omega - \omega'}
(\exp(\beta \omega') +1)     \int_{-\infty}^{+\infty}\exp(i\omega't)
\nonumber\\
<a^{+}_{k+s\sigma}(t)a^{+}_{r-\sigma}(t)a_{r+s-\sigma}(t)
a_{k+p\sigma}a^{+}_{p+q-\sigma}a_{q-\sigma}>
\end{eqnarray}
Further insight is gained if we select the suitable relevant ``trial"
approximation for the correlation function on the r.h.s. of (30). In this
paper we show that the earlier formulations, based on the decoupling or/and
diagrammatic methods can be arrive at from our technique but in a self-
consistent way. Clearly that the choice of the relevant trial
approximation for correlation function in (30)
can be done in many ways. For example, the reasonable
and workable one may be the following ``pair approximation", which is
especially
good for the low density of the quasiparticles:
\begin{eqnarray}
<a^{+}_{k+s\sigma}(t)a^{+}_{r-\sigma}(t)a_{r+s-\sigma}(t)
a_{k+p\sigma}a^{+}_{p+q-\sigma}a_{q-\sigma}>^{ir} \approx \nonumber\\
<a^{+}_{k+p\sigma}(t)a_{k+p\sigma}><a^{+}_{q-\sigma}(t)a_{q-\sigma}>
<a_{p+q-\sigma}(t)a^{+}_{p+q-\sigma}>\nonumber\\
\delta_{k+s, k+p} \delta_{r, q} \delta_{r+s, p+q}
\end{eqnarray}
Using (30) and (31) in (29) we obtain the approximate expression for the
self-energy operator in a self-consistent form (the self-consistency
means that we express approximately the self-energy operator in terms
of the initial GF and, in principle, one can obtain the required solution
by suitable iteration procedure):
\begin{eqnarray}
M_{\sigma}(k, \omega) = \frac{U^2}{N^2} \sum_{pq} \int
\frac{d\omega_{1}d\omega_{2}d\omega_{3}}{\omega + \omega_{1} -
\omega_{2} - \omega_{3}}\nonumber\\
~[n(\omega_{2})n(\omega_{3}) +
n(\omega_{1})(1 - n(\omega_{2}) - n(\omega_{3}))]
g_{p+q-\sigma}(\omega_{1})g_{k+p\sigma}(\omega_{2})g_{q-\sigma}(\omega_{3})
\end{eqnarray}
where we have used the notations
$$g_{k\sigma}(\omega) = -{1 \over \pi}Im G_{k\sigma}(\omega + i\varepsilon);
n(\omega) = [\exp(\beta\omega) + 1]^{-1}$$
The equations (28) and (32) form a closed self-consistent system of
equations
for the single-electron GF for the Hubbard model, but for weakly correlated
limit only. In principle, we may use, on the r.h.s. of (32) any workable
first iteration-step form of the GF and find a solution by repeated
iteration.
It is most convenient to choose as the first iteration step the following
simple one-pole approximation:
\begin{equation}
g_{k\sigma}(\omega) \approx \delta(\omega - \epsilon(k\sigma))
\end{equation}
Then, using (33) in (32), we get for the self-energy an explicit and
simple expression
\begin{equation}
M_{\sigma}(k, \omega) = \frac{U^2}{N^2} \sum_{pq}
\frac{n_{p+q-\sigma}(1 - n_{k+p\sigma} - n_{q-\sigma}) + n_{k+p\sigma}
n_{q-\sigma}}{\omega + \epsilon(p+q\sigma) - \epsilon(k+p\sigma) -
\epsilon(q\sigma)}
\end{equation}
The numerical calculations of the typical behaviour of
real and imaginary parts of the
self-energy
(34) have been performed~\cite{kuz42} for the model density of states
of the FCC lattice. These calculations and many other~\cite{cal43} -
\cite{cal45}
prove that the conventional
one-electron approximation of the band theory is not always a sufficiently
good approximation for transition metals like nickel.
The simple formula (32) derived above for the self-energy operator are
typical
in showing the role of correlation effects in the formation of quasiparticle
spectrum of the Hubbard model. It is instructive to examine other types of
the possible trial solutions for the six-operator correlation function
in the eqn.(30). The approximation which we propose now reflects the
interference between the one-particle branch of the spectrum and the
collective one:
\begin{eqnarray}
<a^{+}_{k+s\sigma}(t)a^{+}_{r-\sigma}(t)a_{r+s-\sigma}(t)
a_{k+p\sigma}a^{+}_{p+q-\sigma}a_{q-\sigma}>^{ir} \approx \nonumber\\
<a^{+}_{k+s\sigma}(t)a_{k+p\sigma}><a^{+}_{r-\sigma}(t)a_{r+s-\sigma}(t)
a^{+}_{p+q-\sigma}a_{q-\sigma}> + \nonumber\\
<a_{r+s-\sigma}(t)a^{+}_{p+q-\sigma}><a^{+}_{k+s\sigma}(t)a^{+}_{r-\sigma}(t
)
a_{k+p\sigma}a_{q-\sigma}> +\nonumber\\
<a^{+}_{r-\sigma}(t)a_{q-\sigma}><a^{+}_{k+s\sigma}(t)a_{r+s-\sigma}(t)
a_{k+p\sigma}a^{+}_{p+q-\sigma}>
\end{eqnarray}
It is visible now that the three contributions in this trial solution
describe the self-energy corrections that take into account the collective
motions of electron density, the spin density and the density of ``doubles",
respectively. The essential feature of this approximation is connected
with the fact that correct calculation of the single-electron quasiparticle
spectra with damping require the suitable incorporating of the influence
of the collective degrees of freedom on the single-particle ones. The most
interesting contribution is related with the spin degrees of freedom
because of correlated system are the magnetic or have very well developed
magnetic fluctuations. We follows the above steps and calculate the
self-energy operator (29) as
\begin{eqnarray}
M_{\sigma}({\vec k}, \omega) = {U^2\over N}\int_{-\infty}^{+\infty}
d\omega_{1}d{\omega}_{2}\frac{1 + N(\omega_{1}) - n(\omega_{2})}
{\omega - \omega_{1} - \omega_{2}}\nonumber\\
\sum_{i, j}\exp[-i{\vec k}({\vec R_{i}} - {\vec R_{j}})]
(-{1 \over \pi}Im <<S^{\pm}_{i} \vert S^{\mp}_{j}>>_{\omega_{1}})\nonumber\\
(-{1 \over \pi}Im<<a_{i-\sigma}\vert a^{+}_{j-\sigma}>>_{\omega_{2}})
\end{eqnarray}
where the following notations have been used:
$$S^{+}_{i} = a^{+}_{i\uparrow}a_{i\downarrow};
S^{-}_{i} = a^{+}_{i\downarrow}a_{i\uparrow}$$
$$N(\omega) = [\exp(\beta \omega) - 1]^{-1}.$$
It is possible to rewrite (37) in a more convenient way now
\begin{eqnarray}
M_{\sigma}(k, \omega) = {U^2\over N} \sum_{q} \int d\omega'
(\cot \frac{\omega - \omega'}{2T} + \tan \frac{\omega'}{2T})\nonumber\\
(- \frac{1}{\pi} Im \chi^{\mp \pm} (k-q, \omega - \omega')g_{q\sigma}
(\omega')).
\end{eqnarray}
The equations (28) and (37) form again another self-consistent system
of equations for the single-particle GF of the Hubbard model. Note, that
both expressions for the self-energy depend on quasimomentum; in other
words the approximate procedure do not broke the momentum conservation
law. It is important, because of the poles $\omega(k, \sigma) =
\epsilon(k, \sigma) -i\Gamma(k)$ of the GF (28) are determined by the
equation
\begin{equation}
\omega - \epsilon(k\sigma) - Re[M_{\sigma}(k, \omega)] -
iIm [M_{\sigma}(k, \omega)] = 0
\end{equation}
It may be shown quite generally that the Luttinger's definition of the
true Fermi surface~\cite{mah7} is valid in the framework of our theory.
It is worthy to note that for electrons in a crystal where there is a band
index, as well as quasimomentum, the definition of the Fermi surface are
a little more complicated then the single-band one. Before the single
particle energies and Fermi surface are known, one must carry out a
diagonalization in the band index.
In order to give a complete picture of the GMFs let us discuss briefly
the interesting question of the correct definition of the so-called
unrestricted Hartree-Fock approximation (UHFA). Recently, this
approximation
has been applied for the single-band Hubbard model (11) for the calculation
of the density of states for $CuO_{2}$ clusters~\cite{lop46}. The following
definition of UHFA has been used:
\begin{equation}
n_{i-\sigma}a_{i\sigma} = <n_{i-\sigma}>a_{i\sigma} -
<a^{+}_{i-\sigma}a_{i\sigma}>a_{i-\sigma}
\end{equation}
Thus, in addition to the standard HF term, the new, the so-called
``spin-flip" terms, are retained. This example clearly show that the nature
of
the mean-fields follows from the essence of the problem and should be
defined in a proper way. It is clear, however, that the definition (39)
broke the rotational symmetry of the Hubbard Hamiltonian. For the single-
band Hubbard Hamiltonian the averaging $<a^{+}_{i-\sigma}a_{i, \sigma}> =
0$ because of the rotational symmetry of the Hubbard model. So, in
Ref. ~\cite{lop46} the effective Hamiltonian $H_{\rm eff}$ has been defined.
We have
analysed in detail the proper definition of the irreducible
GFs which include the ``spin-flip" terms.
The definition (19) must be modified in the following way:
\begin{eqnarray}
^{ir}<<a_{k+p\sigma}a_{p+q-\sigma}a_{p+q-\sigma} \vert a^{+}_{k\sigma}>>_
{\omega} =
<<a_{k+p\sigma}a^{+}_{p+q-\sigma}a_{q-\sigma}>>_{\omega} - \nonumber\\
\delta_{p, 0}<n_{q-\sigma}>G_{k\sigma} - <a_{k+p\sigma}a^{+}_{p+q-\sigma}>
<<a_{q-\sigma} \vert a^{+}_{k\sigma}>>_{\omega}
\end{eqnarray}
From this definition follows that such a type of introduction of the IGF
broaden the initial algebra of the operator and initial set of the GFs.
That means that ``actual" algebra of the operators must include the
spin-flip
terms at the beginning, namely:
$(a_{i\sigma}$, $a^{+}_{i\sigma}$, $n_{i\sigma}$,
$a^{+}_{i\sigma}a_{i-\sigma})$. The corresponding initial GF will have the
form

$$\pmatrix{
<<a_{i\sigma}\vert a^{+}_{j\sigma}>> & <<a_{i\sigma}\vert
a^{+}_{j-\sigma}>> \cr
<<a_{i-\sigma}\vert a^{+}_{j\sigma}>> & <<a_{i-\sigma}\vert
a^{+}_{j-\sigma}>> \cr}$$

In fact, this approximation has
been investigated earlier by Kishore and Jo\-shi~\cite{kis47}. They clearly
pointed out that they assumed that the system is magnetized in $x$ direction
instead of conventional $z$ axis.
\section{Hubbard Model. Strong Correlation}
When studying the electronic quasiparticle spectrum of the strongly
correlated systems, one must take care of at least three facts of major
importance:\\
(i) The ground state is reconstructed radically as compared with the
weakly correlated case. Namely this fact lead to the necessity of the
redefinition of the single-particle states. Due to the strong correlation,
the initial algebra of the operators are transformed into new algebra of
the complicated operators. In principle, in terms of the new operators the
initial Hamiltonian may be rewritten as bilinear form and the
generalized Wick theorem can be formulated~\cite{gau48}, \cite{wes49}. It
is very important to underline, that the transformation to the new
algebra of relevant operators reflect some important internal symmetries
of the problem and nowadays this way of thinking are formulating in elegant
and very powerful technique of the classification of the integrable
models~\cite{cha50}, \cite{hald51} and exactly soluble models
(see also\cite{kuz52}).\\
(ii) The single-electron GF, which describe the dynamical properties,
must have two-pole functional structure, giving in the atomic limit, when
hopping integral tends to zero, the exact two-level atomic solution.\\
(iii) The GMFs have, in general case, a very non-trivial structure. The
GMFs functional cannot be expressed in terms of the functional of the
mean particles density.\\
In this section we consider large, but finite, Coulomb repulsion. The
inspiring ideas of papers~\cite{hub34}, \cite{hal41}, \cite{wes49} where
the problem of the relevant algebra of the operators has been considered,
are central to our consideration here. Following this approach we consider
the new set of relevant operators:
\begin{eqnarray}
d_{i\alpha\sigma} = n^{\alpha}_{i-\sigma}a_{i\sigma}, (\alpha = \pm);
n^{+}_{i\sigma} = n_{i\sigma}, n^{-}_{i\sigma} = (1 - n_{i\sigma});
\nonumber\\
\sum n^{\alpha}_{i\sigma} = 1; n^{\alpha}_{i\sigma}n^{\beta}_{i\sigma} =
\delta_{\alpha\beta}n^{\alpha}_{i\sigma}; \sum_{\alpha} d_{i\alpha\sigma} =
a_{i\sigma}
\end{eqnarray}
The new operators $d_{i\alpha \sigma}$ and $d^{+}_{j\beta \sigma}$ have
complicated commutation rules, namely
$$[d_{i\alpha \sigma}, d^{+}_{j\beta \sigma}]_{+} =
\delta_{ij} \delta_{\alpha \beta}n^{\alpha}_{i-\sigma}$$
The convenience of the new operators follows immediately if one write
down the equation of motion for them
\begin{eqnarray}
[d_{i\alpha \sigma}, H]_{-} = E_{\alpha}d_{i\alpha \sigma} +
\sum_{ij}t_{ij}(n^{\alpha}_{i-\sigma}a_{j\sigma} + \alpha a_{i\sigma}
b_{ij-\sigma})\nonumber\\
b_{ij\sigma} = (a^{+}_{i\sigma}a_{j\sigma} - a^{+}_{j\sigma}a_{i\sigma}).
\end{eqnarray}
It is possible to interpret~\cite{hub18}, \cite{hub34} both contribution
in this equation as {\it alloy analogy} and {\it resonance broadening
correction}.
Let us consider the single-particle GF (12) in the Wannier basis. Using the
new operator algebra it is possible to rewrite identically GF (12) in the
following way
\begin{equation}
G_{ij \sigma}(\omega) = \sum_{\alpha \beta}<<d_{i\alpha \sigma} \vert
d^{+}_{j\beta \sigma}>>_{\omega} = \sum_{\alpha \beta} F^{\alpha \beta}_{
ij\sigma}(\omega)
\end{equation}
The equation of motion for the auxiliary matrix GF
\begin{equation}
F^{\alpha \beta}_{ij\sigma}(\omega) = \pmatrix{<<d_{i+\sigma} \vert
d^{+}_{j+\sigma}>>_{\omega}&<<d_{i+\sigma} \vert d^{+}_{j-\sigma}>>_{\omega}
 \cr
<<d_{i-\sigma} \vert d^{+}_{j+\sigma}>>_{\omega}&<<d_{i-\sigma} \vert
d^{+}_{j-\sigma}>>_{\omega}\cr}
\end{equation}
have the following form
\begin{equation}
({\bf E}{\bf F}_{ij\sigma}(\omega) -{\bf I} \delta_{ij})_{\alpha \beta} =
\sum_{l\not= i}t_{il}<<n^{\alpha}_{i-\sigma}a_{l\sigma} + \alpha a_{i\sigma}
b_{il-\sigma} \vert d^{+}_{j\beta \sigma}>>_{\omega}
\end{equation}
Where the following matrix notations have been used
\begin{equation}
{\bf E} = \pmatrix{(\omega- E_{+})&0\cr
0&(\omega - E_{-})\cr} ;
{\bf I} = \pmatrix{n^{+}_{-\sigma}&0\cr
0&n^{-}_{-\sigma} \cr}.
\end{equation}
In accordance with the general method of Section 2 we introduce by
definition
the matrix IGF:
\begin{eqnarray}
{\bf D}^{ir}_{il, j}(\omega) = \pmatrix{<<Z_{11}\vert d^{+}_{j+\sigma}>>_
{\omega}&<<Z_{12}\vert d^{+}_{j-\sigma}>>_{\omega}\cr
<<Z_{21}\vert d^{+}_{j+\sigma}>>_{\omega}&<<Z_{22}\vert d^{+}_{j-\sigma}>>_
{\omega}\cr} -\nonumber\\
\sum _{\alpha'}({A^{+\alpha'}_{il}\brack A^{-\alpha'}_{il}}[F^{\alpha'+}_{
ij\sigma} \ F^{\alpha'-}_{ij\sigma}] - {B^{+\alpha'}_{li}\brack
B^{-\alpha'}_{
li}}[F^{\alpha'+}_{lj\sigma} \ F^{\alpha'-}_{lj\sigma}])
\end{eqnarray}
Here the notations have been used:
$$Z_{11} = Z_{12} = n^{+}_{i-\sigma}a_{l\sigma} + a_{i\sigma}b_{il-\sigma};
\
Z_{21} = Z_{22} = n^{-}_{i-\sigma}a_{l\sigma} - a_{i\sigma}b_{il-\sigma}$$
It is worth to underline that the definition (47) are in heart of the whole
our approach to description of the strong correlation in the Hubbard model.
The coefficients A and B are determined from the constraint (4), namely
\begin{equation}
<[({\bf D}^{ir}_{il, j})_{\alpha \beta}, d^{+}_{j\beta \sigma}]_{+}> = 0
\end{equation}
After some algebra we obtain from (48) ($i \not= j$)
\begin{eqnarray}
[A_{il}]_{\alpha \beta} = \alpha(<d^{+}_{i\beta-\sigma}a_{l-\sigma}> +
<d_{i-\beta-\sigma}a^{+}_{l-\sigma}>)(n^{\beta}_{-\sigma})^{-1}\nonumber\\
~[B_{li}]_{\alpha \beta} = [<n^{\beta}_{l-\sigma}n^{\alpha}_{i-\sigma}> +
\alpha \beta(<a_{i\sigma}a^{+}_{i-\sigma}a_{l-\sigma}a^{+}_{l\sigma}> -
\nonumber\\
<a_{i\sigma}a_{i-\sigma}a^{+}_{l-\sigma}a^{+}_{l\sigma}>)](n^{\beta}_{
-\sigma})^{-1}
\end{eqnarray}
As previously, we introduce now the GMF GF ${\bf F}^{0}_{ij\sigma}$ in
analogy
with (6), however, as it is clear from (47), the actual definition of the
GMF GF is very non-trivial. After the Fourier transformation we get
\begin{equation}
\pmatrix{
F^{0++}_{k\sigma}&F^{0+-}_{k\sigma}\cr
F^{0-+}_{k\sigma}&F^{0--}_{k\sigma}\cr} =
\frac 1 {ab - cd} %
\pmatrix{
 n^{+}_{-\sigma}b & n^{-}_{-\sigma}d \cr %
 n^{+}_{-\sigma}c & n^{-}_{-\sigma}a }   %
% \frac {n^{+}_{-\sigma}b}{ab - cd}&\frac {n^{-}_{-\sigma}d}{ab - cd}\cr
% \frac {n^{+}_{-\sigma}c}{ab - cd}&\frac{n^{-}_{-\sigma}a}{ab - cd}\cr}
\end{equation}
The coefficients $a$, $b$, $c$, $d$ are equal to
\begin {eqnarray}
{a\atop b} = (\omega - E_{\pm} - N^{-1}\sum_{p}\epsilon(p)(A^{\pm \pm}(-p) -
B^{\pm \pm}(p - q)))\nonumber\\
{c\atop d} = N^{-1}\sum_{p}\epsilon(p)(A^{\mp \pm}(-p) -
B^{\mp \pm}(p - q))
\end{eqnarray}
Then, using the definition (43) we find the final expression for the GMF GF
\begin{equation}
G^{MF}_{\sigma}(k, \omega) =
\frac{\omega - (n^{+}_{-\sigma}E_{-} + n^{-}_{-\sigma}E_{+}) - \lambda(k)}{
(\omega -E_{+} - n^{-}_{-\sigma}\lambda_{1}(k))(\omega - E_{-} -
 n^{+}_{-\sigma}
\lambda_{2}(k)) -
n^{-}_{-\sigma}n^{+}_{-\sigma}\lambda_{3}(k)\lambda_{4}(k)}
\end{equation}
Here we have introduced the following notations:
\begin{eqnarray}
{\lambda_{1}(k)\atop \lambda_{2}(k)} = \frac{1}{n^{\mp}_{-\sigma}}\sum_{p}
\epsilon(p)(A^{\pm \pm}(-p) - B^{\pm \pm}(p - k))\\
{\lambda_{3}(k)\atop \lambda_{4}(k)} = \frac{1}{n^{\mp}_{-\sigma}}\sum_{p}
\epsilon(p)(A^{\pm \mp}(-p) - B^{\pm \mp}(p - k))\\
\lambda(k) = (n^{-}_{-\sigma})^{2}(\lambda_{1} + \lambda_{3}) +
(n^{+}_{-\sigma})^{2}(\lambda_{2} + \lambda_{4}) \nonumber
\end{eqnarray}
From the equation (52) it is obvious that our two-pole solution is more
general
than ``Hubbard III" \cite{hub34} and Roth\cite{roth36} solutions. Our
solution
has the correct non-local structure, taking into account the non-diagonal
scattering matrix elements more accurately. Those matrix elements describe
the virtual ``recombination" processes and reflect the extremely complicated
structure of the single-particle state, which virtually include a great
number
of intermediate scattering processes (c.f. interesting analysis in
Ref. ~\cite{clar53}). \\
The spectrum of the mean-field quasiparticle excitations follows from the
poles of the GF (52) and consist of two branches
\begin{equation}
\omega{1\atop 2}(k) = 1/2[(E_{+} - E_{-} + a_{1} + b_{1}) \pm \sqrt
{(E_{+} +E_{-} - a_{1} - b_{1})^{2} -4cd}]
\end{equation}
where $a_{1}(b) = \omega - E_{\pm} -a(b)$. Thus the Spectral Intensity
$A_{k\sigma}(\omega)$ of the GF (52) consist of two peaks, which separated
by the distance
\begin{equation}
\omega_{1} - \omega_{2} =\sqrt{(U - a_{1} - b_{1})^{2} - cd} \approx
U( 1 - \frac{a_{1} - b_{1}}{U}) + O(\gamma)
\end{equation}
For the deeper insight into the functional structure of the solution
(52) and to compare with the other solutions we rewrite the (50) in the
following form
\begin{equation}
{\bf F}^{0}_{k\sigma}(\omega) = \pmatrix{
({a\over n^{+}_{-\sigma}} - {db^{-1}c\over n^{+}_{-\sigma}})^{-1}&{d
\over a}({b\over n^{-}_{-\sigma}} - {da^{-1}c\over n^{-}_{-\sigma}})^{-1}\cr
{c\over b}({a\over n^{+}_{-\sigma}} - {db^{-1}c\over n^{+}_{-\sigma}})^{-1}&
({b\over n^{-}_{-\sigma}} - {db^{-1}c\over n^{-}_{-\sigma}})^{-1}\cr}
\end{equation}
from which we obtain for the $G^{MF}_{\sigma}(k, \omega)$
\begin{eqnarray}
G^{MF}_{\sigma}(k, \omega) = \frac {n^{+}_{-\sigma}(1 + cb^{-1})}{a -
db^{-1}c} + \frac {n^{-}_{-\sigma}(1 + da^{-1})}{b - ca^{-1}d} \approx
\nonumber\\
\frac {n^{-}_{-\sigma}}{\omega - E_{-} - n^{+}_{-\sigma}W^{-}_{-\sigma}(k)}
+
\frac {n^{+}_{-\sigma}}{\omega - E_{+} - n^{-}_{-\sigma}W^{+}_{-\sigma}(k)}
\end{eqnarray}
where
\begin{eqnarray}
n^{+}_{-\sigma}n^{-}_{-\sigma}W^{\pm}_{-\sigma}(k) = N^{-1}\sum_{ij} t_{ij}
\exp(-ik(R_{i} -R_{j}))\nonumber\\
\left ((<a^{+}_{i-\sigma}n^{\pm}_{i\sigma}a_{j-\sigma}> + <a_{i-\sigma}
n^{\mp}_{i\sigma}a^{+}_{j-\sigma}>) + \right . \nonumber\\
\left . (<n^{\pm}_{j-\sigma}n^{\pm}_{i-\sigma}> +
<a_{i\sigma}a^{+}_{i-\sigma}
a_{j-\sigma}a^{+}_{j\sigma}> - <a_{i\sigma}a_{i-\sigma}a^{+}_{j-\sigma}
a^{+}_{j\sigma}>) \right)
\end{eqnarray}
are the shifts for the upper and lower splitted subbands due to the elastic
scattering of the carriers in the Generalized Mean Field. Namely $W^{\pm}$
are the functionals of the GMF. The most important feature of the present
solution of the strongly correlated Hubbard model is a very nontrivial
structure of the mean-field renormalizations (59), which is crucial to
understanding the physics of strongly correlated systems. It is important
to emphasize that namely this complicated form of the GMF are only
relevant to the essence of the physics under consideration. The attempts to
reduce the functional of the GMF to the simpler functional of the average
density of electrons are incorrect namely from the point of view of the
real nature of the physics of HCS. This physics clearly show that the
mean-field renormalizations cannot be expressed as a functional of the
electron mean density. To explain this statement let us derive the
``Hubbard I" solution~\cite{hub18} from our GMF solution (52). If we
approximate (59) as
\begin{equation}
n^{+}_{-\sigma}n^{-}_{-\sigma}W^{\pm}(k) \approx N^{-1}\sum_{ij}t_{ij}
{\exp(-ik(R_{i} - R_{j}))}<n^{\pm}_{j-\sigma}n^{\pm}_{i-\sigma}>
\end{equation}
and makes the additional approximation, namely $$<n_{j-\sigma}n_{i-\sigma}>
\approx n^{2}_{-\sigma}$$ then solution (52) goes over into the
``Hubbard I" solution
\begin{equation}
G^{0}_{\sigma}(k, \omega) \approx \frac {n_{-\sigma}}{\omega - U -
\epsilon(k)n_{-\sigma}} + \frac {1 - n_{-\sigma}}{\omega - \epsilon(k)
(1 - n_{-\sigma})}
\end{equation}
This solution, as it is well known, is unrealistic from the many points
of view. \\

As regards to our solution (52), the second important aspect is that the
parameters $\lambda_{i}(k)$ do not depend on frequency, i.e. depends
essentially on the elastic scattering processes. Such a dependence on
frequency
arises due to inelastic scattering processes which are contained in our
self-energy operator and we proceed now with the derivation of the
explicit expression for it. \\
To calculate the high-order GF on the r.h.s. of (45) we should use the
second time variable ($t'$) differentiation of it again. If one introduces
irreducible parts for the right-hand-side operators by analogy with
expression (47), the equation of motion (45) can be rewritten exactly in
the following form
\begin{equation}
{\bf F}_{k\sigma}(\omega) = {\bf F}^{0}_{k\sigma}(\omega) +
{\bf F}^{0}_{k\sigma}(\omega){\bf P}_{k\sigma}(\omega){\bf F}^{0}_{k\sigma}
(\omega)
\end{equation}
Here the scattering operator $P$ (8) has the form
\begin{equation}
{\bf P}_{q\sigma}(\omega) = {\bf I}^{-1}[\sum_{lm}t_{il}t_{mj}
<<{\bf D}^{ir}_{il, j}|{\bf D}^{ir +}_{i, mj}>>_{\omega}]_{q}{\bf I}^{-1}
\end{equation}
In accordance with the definition (9) we write down the Dyson equation
\begin{equation}
{\bf F} = {\bf F}^{0} + {\bf F}^{0}{\bf M}{\bf F}
\end{equation}
The self-energy operator $M$ is defined by Eq. (9). Let us note again that
the self-energy corrections, according to (10), contribute to the full
GF as an additional terms. This is an essential advantage in comparison
with the ``Hubbard III" solution and other two-pole solutions. For the full
GF we find, using the formal solution of Dyson equation
\begin{eqnarray}
G_{\sigma}(k, \omega) = \left ( {1\over n^{+}_{-\sigma}}(a - n^{+}_{-\sigma}
M^{++}_{\sigma}(k, \omega)) + {1\over n^{-}_{-\sigma}}(b - n^{-}_{-\sigma}
M^{--}_{\sigma}(k, \omega)) \right . \nonumber\\
\left . + {1\over n^{+}_{-\sigma}}(d + n^{+}_{-\sigma}
M^{+-}_{\sigma}(k, \omega)) + {1\over n^{-}_{-\sigma}}(c + n^{-}_{-\sigma}
M^{-+}_{\sigma}(k, \omega))\right )\nonumber\\
~[{\rm det}\left ( (F^{0}_{k\sigma}
(\omega))^{-1} - M_{\sigma}(k, \omega)\right )]^{-1}
\end{eqnarray}
After some algebra we can rewrite this expression in the following form,
which is essentially new and, in a certain sense, are the central result
of the present theory
\begin{equation}
G = \frac {\omega - (n^{+}E_{-} + n^{-}E_{+}) - L}{(\omega -E_{+}
-n^{-}L_{1})
(\omega - E_{-} -n^{+}L_{2}) - n^{-}n^{+}L_{3}L_{4}}
\end{equation}
where
\begin{eqnarray}
L_{1}(k, \omega) = \lambda_{1}(k) - {n^{+}_{-\sigma}\over n^{-}_{-\sigma}}
M^{++}_{\sigma}(k, \omega);\nonumber\\
L_{2}(k, \omega) = \lambda_{2}(k) - {n^{-}_{-\sigma}\over n^{+}_{-\sigma}}
M^{--}_{\sigma}(k, \omega);\nonumber\\
L_{3}(k, \omega) = \lambda_{3}(k) + {n^{-}_{-\sigma}\over n^{+}_{-\sigma}}
M^{+-}_{\sigma}(k, \omega);\nonumber\\
L_{4}(k, \omega) = \lambda_{4}(k) + {n^{+}_{-\sigma}\over n^{-}_{-\sigma}}
M^{-+}_{\sigma}(k, \omega);\nonumber\\
L(k, \omega) = \lambda(k) + n^{+}_{-\sigma}n^{-}_{-\sigma}(M^{++} + M^{--} -
M^{-+} - M^{+-})
\end{eqnarray}
Thus, now we have to find the explicit expressions for the elements of the
self-energy matrix M. To proceed we should use the spectral theorem again,
as in Eq. (30), to express the GF in terms of correlation functions
\begin{equation}
M^{\alpha, \beta}_{\sigma}(k, \omega) \sim <D^{ir +}_{mj, \beta}(t)
D^{ir}_{il, \alpha}>
\end{equation}
For the approximate calculation of the self-energy we propose to use
the following trial solution
\begin{eqnarray}
<D^{ir +}(t)D^{ir}> \approx <a^{+}_{m\sigma}(t)a_{l\sigma}><n^{\beta}_
{j-\sigma}(t)n^{\alpha}_{i-\sigma}>\nonumber\\
 + <a^{+}_{m\sigma}(t)n^{\alpha}_{i-\sigma}>
<n^{\beta}_{j-\sigma}(t)a_{l\sigma}> +\beta
<b^{+}_{mj-\sigma}(t)a_{l\sigma}>
<a^{+}_{j\sigma}(t)n^{\alpha}_{i-\sigma}>\nonumber\\
 + \beta <b^{+}_{mj-\sigma}(t)
n^{\alpha}_{i-\sigma}><a^{+}_{j\sigma}(t)a_{l\sigma}> + \alpha
 <a^{+}_{m\sigma}
(t)a_{i\sigma}><n^{\beta}_{j-\sigma}(t)b_{il-\sigma}>\nonumber\\
 + \alpha <a^{+}_{m\sigma}
(t)b_{il-\sigma}><n^{\beta}_{j-\sigma}(t)b_{il-\sigma}>\nonumber\\
 + \alpha\beta
 <b^{+}_{mj-\sigma}(t)a_{i\sigma}><a^{+}_{j\sigma}(t)b_{il-\sigma}>
\nonumber\\
 +
\alpha \beta <b^{+}_{mj-\sigma}(t)b_{il-\sigma}>
<a^{+}_{j\sigma}(t)a_{i\sigma}>
\end{eqnarray}
It is quite natural to interpret the contributions in this expression in
terms of scattering, resonance-broadening and interference corrections of
different types. For example, let us consider the simplest approximation.
For this aim we retain the first contribution in (69)
\begin{eqnarray}
[{\bf I}{\bf M}{\bf I}]_{\alpha \beta} = \int_{-\infty}^{+\infty}
{d\omega'\over \omega - \omega'}(\exp(\beta \omega') + 1)\nonumber\\
\int_{-\infty}^{+\infty}{dt\over 2\pi}\exp(i\omega't)N^{-1}\sum_{ijlm}
exp(-ik(R_{i}-R_{j}))t_{il}t_{mj}\nonumber\\
\int d\omega_{1}n(\omega_{1})
\exp(i\omega_{1}t)g_{ml\sigma}(\omega_{1})\left (-{1\over \pi}Im
K^{\alpha \beta}_{ij}(\omega_{1} - \omega')\right ).
\end{eqnarray}
Equations (70) and (64) are the self-consistent system of equations for
the single-particle Green's function. For a simple estimation, for the
calculation of the self-energy (70) it is possible to use any initial
relevant approximation of the two-pole structure. As an example we take the
expression (61). We then obtain
\begin{eqnarray}
[{\bf I}{\bf M}{\bf I}]_{\alpha \beta} \approx \sum_{q}|\epsilon(k - q)|^{2}
K^{\alpha \beta}_{q}\nonumber\\
~[\frac{n_{-\sigma}}{\omega - U - \epsilon(k-q)n_{-\sigma}} +
\frac{1 - n_{-\sigma}}{\omega - \epsilon(k-q)(1 - n_{-\sigma})}]
\end{eqnarray}
On the basis of the self-energy operator (71) we can explicitly find
the energy shift and damping due to inelastic scattering of the
 quasiparticles, which is a great advantage of the present approach. It
is clear from the present consideration that for the systematic construction
of the approximate solutions we need to calculate the collective
correlation functions of the electron density and spin density and the
density of doubles, but this problem must be considered separately.
\section{Correlations in Random Hubbard Model}
In this chapter we shall apply IGF method for consideration of the
electron-electron correlations in the presence of disorder to demonstrate
the
advantage of our approach. The treatment of the electron motion in
substitutionally disordered disordered $A_{x}B_{1-x}$ transition metal
alloys is based upon certain generalization of Hubbard model, including
random diagonal and off-diagonal elements caused by substitutional
disorder in the binary alloy. The electron-electron interaction play an
important role for various aspects of behaviour in alloys, e.g. in the
weak localisation in Ti-Al alloys\cite{lin54} (for recent review
see~\cite{kir55}).
There are certain aspects
of the High-$T_c$ superconductivity where disorder play a role and
recently  it have been discussed
in  papers\cite{baum56}, \cite{ben57}, where the
distribution of magnetic molecular fields has been treated within the
single-site Coherent Potrntial Approximation(CPA)~\cite{sov58}. The CPA
has been refined and developed in many papers (e.g. \cite{argy59},
\cite{zin60}) and till now\cite{datt61} are the most popular
approximation for theoretical
studying of alloys. But the simultaneous effect of disorder and electron-
electron inelastic scattering has been considered for some limited
cases only\cite{abit62},\cite{bre63} and not within the
self-consistent scheme.
Let us consider the Hubbard model Hamiltonian on a given configuration
of alloy $(\nu)$
\begin{equation}
H^{(\nu)} = H^{(\nu)}_{1} + H^{(\nu)}_{2}
\end{equation}
where
\begin{eqnarray}
H^{(\nu)}_{1} = \sum_{i\sigma}\varepsilon^{\nu}_{i}n_{i\sigma} +
\sum_{ij\sigma}t^{\nu\mu}_{ij}a^{+}_{i\sigma}a_{j\sigma}\nonumber\\
H^{(\nu)}_{2} = {1\over 2}\sum_{i\sigma}U^{\nu}_{i}n_{i\sigma}n_{i-\sigma}
\end{eqnarray}
Contrary to the periodic model (11), the atomic level energy
$\varepsilon^{\nu}_{i}$, the hopping integrals $t^{\nu\mu}_{ij}$ as well as
the intraatomic
Coulomb repulsion $U^{\nu}_{i}$ here are the random variables, which take
the values $\varepsilon^{\nu}$, $t^{\nu\mu}$ and $U^{\nu}$, respectively;
the superscript $\nu(\mu)$ refers to the atomic species ($\nu, \mu = A, B$)
located on site i(j). The nearest-neighbour hopping integrals are included
only. \\
To unify the IGF method and CPA into completely self-consistent scheme
let us consider the single-electron GF (17) $G_{ij\sigma}$
 in the Wannier representation for
a given configuration $(\nu)$. The corresponding equation of motion has
the form(for brevity we shall omit the superscript $(\nu)$ where its
presence is clear)
\begin{eqnarray}
(\omega - \varepsilon_{i})<<a_{i\sigma}|a^{+}_{j\sigma}>>_{\omega} =
\delta_{ij} + \sum_{n}t_{in}<<a_{n\sigma}|a^{+}_{j\sigma}>>_{\omega}
\nonumber\\
+ U_{i}<<n_{i-\sigma}a_{i\sigma}|a^{+}_{j\sigma}>>_{\omega}
\end{eqnarray}
In the present paper, for brevity, we will confine ourselves by the weak
correlation case and the diagonal disorder only. The generalization for
the case of strong correlation or off-diagonal disorder is straightforward,
but its length considerations preclude us from discussing at this time. \\
Using the definition (3), we define the IGF for a given (fixed) configuration
of atoms in an alloy as follows
\begin{equation}
^{ir}<<n_{i-\sigma}a_{i\sigma}|a^{+}_{j\sigma}>> =
<<n_{i-\sigma}a_{i\sigma}|
a^{+}_{j\sigma}>> - <n_{i-\sigma}><<a_{i\sigma}|a^{+}_{j\sigma}>>
\end{equation}
This time, contrary to (19), because of lack of translational invariance
we must take into account the site dependence of $<n_{i-\sigma}>$. Then we
rewrite the equation of motion (76)in the following form
\begin{eqnarray}
\sum_{n}[(\omega - \varepsilon_{i} - U_{i}<n_{i-\sigma}>)\delta_{ij} -
t_{in}]<<a_{n\sigma}|a^{+}_{j\sigma}>>_{\omega} = \nonumber\\
\delta_{ij} +U_{i}(
^{ir}<<n_{i-\sigma}a_{i\sigma}|a^{+}_{j\sigma}>>_{\omega})
\end{eqnarray}
In accordance with the general method of Section 2, we find then the
Dyson equation for a given configuration $(\nu)$
\begin{equation}
G_{ij\sigma}(\omega) = G^{0}_{ij\sigma}(\omega) + \sum_{mn}G^{0}_{im\sigma}
(\omega)M_{mn\sigma}(\omega)G_{nj\sigma}(\omega)
\end{equation}
The GMF GF $G^{0}_{ij\sigma}$ and the self-energy operator $M$ are defined
as
\begin{eqnarray}
\sum_{m}H_{im\sigma}G^{0}_{mj\sigma}(\omega) = \delta_{ij}\nonumber\\
P_{mn\sigma} = M_{mn\sigma} + \sum_{ij}M_{mi\sigma}G^{0}_{ij\sigma}
P_{jn\sigma}\nonumber\\
H_{im\sigma} = (\omega - \varepsilon_{i} - U_{i}<n_{i-\sigma}>)\delta_{im} -
t_{im}\nonumber\\
P_{mn\sigma}(\omega) = U_{m}( ^{ir}<<n_{m-\sigma}a_{m\sigma}|n_{n-\sigma}
a^{+}_{n\sigma}>>^{ir}_{\omega}) U_{n}
\end{eqnarray}
In order to calculate the self-energy operator $M$ self-consistently we have
to express it approximately by the lower-order GFs. Employing the same
pair approximation as (31) (now in Wannier representation) and the same
procedure of calculations we arrive at the following expression for $M$ for
a given configuration $(\nu)$
\begin{eqnarray}
M^{(\nu)}_{mn\sigma}(\omega) = U_{m}U_{n}{1\over 2\pi^{4}}\int R(\omega_{1},
\omega_{2}, \omega_{3})\nonumber\\
ImG^{(\nu)}_{nm-\sigma}(\omega_{1})ImG^{(\nu)}_{
mn-\sigma}(\omega_{2})ImG^{(\nu)}_{mn\sigma}(\omega_{3});\nonumber\\
R = \frac{d\omega_{1}d\omega_{2}d\omega_{3}}{\omega + \omega_{1} -
\omega_{2} - \omega_{3}}\frac{(1 - n(\omega_{1}))n(\omega_{2})n(
\omega_{3})}{n(\omega_{2} + \omega_{3} - \omega_{1})}
\end{eqnarray}
As we have mentioned previously, all the calculations just presented have
been done for a given configuration of atoms in alloy. All the quantities
in our theory (G, $G^{0}$, P, M) depends on the whole configuration of the
alloy. To obtain a theory of a real macroscopic sample, we have to average
over various configurations of atoms in the sample. The configurational
averaging cannot be exactly made for a macroscopic sample. Hence we must
resort to an additional approximation. It is obvious that self-energy $M$
is in turn the functional of $G$, namely $M = M[G]$. If the process of
taking
configurational averaging is denoted by $\bar{G}$, than we have
$$\bar{G} = \bar{G}^{0} + \overline{G^{0}MG}$$
Few words are now appropriate for the description of general possibilities.
The calculations of $\bar{G}^{0}$ can be performed with the help of any
relevant
available scheme. In the present work, for the sake of simplicity, we
choose the single-site CPA\cite{sov58}, namely we take
\begin{equation}
\bar{G}^{0}_{mn\sigma}(\omega) = N^{-1}\sum_{k}\frac{\exp(ik(R_{m}-R_{n}))}
{\omega - \Sigma^{\sigma}(\omega) -\epsilon(k)}
\end{equation}
Here $\epsilon(k) = \sum^{z}_{n=1}t_{n, 0}\exp(ikR_{n})$, z is the number
of nearest neigbours of the site 0, and the Coherent potential $\Sigma^{
\sigma}(\omega)$ is the solution of the CPA self-consistency equations.
For the $A_{x}B_{1-x}$ these read
\begin{eqnarray}
\Sigma^{\sigma}(\omega) = x\varepsilon^{\sigma}_{A} + (1-x)\varepsilon^{
\sigma}_{B} - (\varepsilon^{\sigma}_{A} - \Sigma^{\sigma})F^{\sigma}
(\omega, \Sigma^{\sigma})(\varepsilon^{\sigma}_{B} - \Sigma^{\sigma});
\nonumber\\
F^{\sigma}(\omega, \Sigma^{\sigma}) = \bar{G}^{0}_{mm\sigma}(\omega)
\end{eqnarray}
Now, let us return to the calculation of the configurationally averaged
total GF $\bar{G}$. To perform the remaining averaging in the Dyson equation
we use the approximation
$$\overline{G^{0}MG} \approx \bar{G}^{0}\bar{M}\bar{G}$$
The calculation of $\bar{M}$ requires further averaging of the product of
matrices. We again use the prescription of the factorisability there,
namely
$$\bar{M} \approx \overline{(U_{m}U_{n})}~ \overline{(ImG)}~
 \overline{(ImG)}~ \overline{(ImG)}$$
However, the quantities $\overline{U_{m}U_{n}}$ entering into $\bar{M}$
are averaged here according to
\begin{eqnarray}
 \overline{U_{m}U_{n}} = U_{2} + (U_{1} - U_{2})\delta_{mn}\nonumber\\
 U_{1} = x^{2}U^{2}_{A} + 2x(1-x)U_{A}U_{B} + (1-x)^{2}U^{2}_{B}\nonumber\\
 U_{2} = xU^{2}_{A} + (1-x)U^{2}_{B}
\end{eqnarray}
The averaged value for the self-energy is
\begin{eqnarray}
\bar{M}_{mn\sigma}(\omega) = {U_{2}\over 2\pi^{4}}\int R(\omega_{1},
\omega_{2},
\omega_{3})Im\bar{G}_{nm-\sigma}(\omega_{1})Im\bar{G}_{mn-\sigma}(
\omega_{2})Im\bar{G}_{mn\sigma}(\omega_{3}) +\nonumber\\
\frac{U_{1}-U_{2}}{2\pi^{4}}\delta_{mn}\int R(\omega_{1}, \omega_{2},
\omega_{3})Im\bar{G}_{nm-\sigma}(\omega_{1})Im\bar{G}_{mn-\sigma}(\omega_{2}
)
Im\bar{G}_{mn\sigma}(\omega_{3})
\end{eqnarray}
The averaged quantities are periodic, so we can introduce the Fourier
transform of them, i.e.
$$\bar{M}_{mn\sigma}(\omega) = N^{-1}\sum_{k}\bar{M}_{\sigma}(k, \omega)
\exp(ik(R_{m} - R_{n}))$$
and similar formulae for $\bar{G}$ and $\bar{G}^{0}$. Performing the
configurational
averaging of Dyson equation and Fourier transforming the resulting
expressions according to the above rules, we obtain
\begin{equation}
\bar{G}_{k\sigma}(\omega) = (\omega - \epsilon(k) - \Sigma^{\sigma}(\omega)
-
\bar{M}_{\sigma}(k, \omega))^{-1}
\end{equation}
where
\begin{eqnarray}
\bar{M}_{\sigma}(k, \omega) = {1\over 2\pi^{4}}\sum_{pq} \int R(\omega_{1},
\omega_{2}, \omega_{3})N^{-2}Im\bar{G}_{p-q-\sigma}(\omega_{1})
Im\bar{G}_{q-\sigma}(\omega_{2})\nonumber\\
~[U_{2}Im\bar{G}_{k+p\sigma}(\omega_{3}) + {(U_{1}-U_{2})\over N}
\sum_{g}Im\bar{G}_{k+p-g}(\omega_{3})]
\end{eqnarray}
The simplest way to obtain the explicit solution for the
self-energy$\bar{M}$
is to start with suitable initial trial solution as it was done for the
periodic case (33). For the disordered system, it is reasonable to use
as the first iteration approximation the so-called Virtual Crystal
Approximation(VCA):
$${-1\over \pi}Im\bar{G}^{VCA}_{k\sigma}(\omega +i\epsilon) \approx
\delta(\omega - E^{\sigma}_{k})$$
where for the binary alloy $A_{x}B_{1-x}$ this approximation read
$$\bar{V} = xV^{A} + (1-x)V^{B};\quad E^{\sigma}_{k} =\bar{\varepsilon}
^{\sigma}_{i} + \epsilon(k);$$\\
$$\bar{\varepsilon}^{\sigma}_{i} = x\varepsilon^{\sigma}_{A} + (1-x)
\varepsilon^{\sigma}_{B}$$
Note, that the using of VCA here is by no means the solution of the
correlation problem in VCA. It is only the using the VCA for the
parametrisation of the problem, to start with VCA input parameters.
After the integration of (83) the final result for the self-energy
is
\begin{eqnarray}
\bar{M}_{\sigma}(k, \omega) = {U^{2}\over
N^{2}}\sum_{pq}\frac{n(E^{-\sigma}_{
p+q})[1 - n(E^{-\sigma}_{q}) - n(E^{\sigma}_{k+p})] + n(E^{\sigma}_{k+p})
n(E^{-\sigma}_{q})}{\omega + E^{-\sigma}_{p+q} - E^{-\sigma}_{q} -
E^{\sigma}_{k+p}} + \nonumber\\
{(U_{1}-U_{2})\over N^{3}}\sum_{pqg}\frac{n(E^{-\sigma}_{p+q})
[1 - n(E^{-\sigma}_{q}) - n(E^{\sigma}_{k+p-g})] + n(E^{\sigma}_{k+p-g})
n(E^{-\sigma}_{q})}{\omega + E^{-\sigma}_{p+q} - E^{-\sigma}_{q} -
E^{\sigma}_{k+p-g}}
\end{eqnarray}
It must be emphasized that the equations (84) - (85) give the general
microscopic self-consistent description of inelastic electron-electron
scattering in alloy in the spirit of the CPA. We take into account the
randomness not only through the parameters of the Hamiltonian but also
in a self-consistent way through the configurational dependence of the
self-energy operator.
\section{Electron-Lattice Interaction and MTBA}
In order to understand quantitatively the electrical, thermal and
superconducting properties of metals and their alloys one needs a
proper description an electron-lattice interaction
too~\cite{kuz27},~\cite{kuz64a}.  A systematic, self-consistent
simultaneous treatment of the electron-electron and electron-phonon
interaction plays an important role in recent studies of strongly
correlated systems. It was argued from the different points of view
that in order to understand quantitatively the phenomenon of
high-temperature superconductivity one needs a proper involving of
electron-phonon interaction, too~\cite{car65a}-~\cite{zie76b}.
A lot of theoretical
searches for the relevant mechanism of high temperature
superconductivity deal with the strong electron- phonon
models.  This mechanism
is certainly valuable for bismuthate ceramics~\cite{kuz67} and for
fullerens. Recently~\cite{cav68} a new family of quaternary
intermetallic $LuNi_{2}B_{2}C$ compounds has been shown to display
superconductivity with Tc = 16.6 K for $LuNi_{2}B_{2}C$, which, besides
that studies of their physical properties are still in the early
stages, suggest that electron-phonon coupling is responsible for the
superconductivity.\\ The natural approach for the description of
superconductivity in such type of compounds is MTBA~\cite{kuz27},
\cite{kuz64a}. The papers ~\cite{kuz27}, \cite{wys28},\cite{wys29}
contain a self-consistent microscopic theory of the normal and
superconducting properties of transition metals and strongly disordered
binary alloys in the framework of Hubbard Model (11) and random Hubbard
model (73). It is worthy to emphasize that in paper~\cite{wys29} a very
detailed microscopic theory of strong coupling superconductivity in
highly disordered transition metals alloys has been developed on the
basis of IGF method within MTBA reformulated approach~\cite{wys28}. The
Eliashberg-type strong coupling equations for highly disordered alloys
has been derived. It was shown that the electron-phonon Spectral
Function in alloy is modified strongly.  An interesting
discussion~\cite{kim70}, \cite{abr71}, \cite{kim72}
clarified many uncertainties in this
important issue (c.f.~\cite{arb72a}-~\cite{ban74a}).
\section{Other Applications of the IGFs Method} Another
important application of IGF method is related with the investigation
of non-local correlations and quasiparticle interactions in Anderson
model~\cite{kuz24}, \cite{kuz25}. A comparative study of real-many body
dynamics of single-impurity, two-impurity and periodic Anderson model,
especially for strong but finite Coulomb correlation, when perturbation
expansion in $U$ does not work (c.f.~\cite{san73}) has permitted  to
characterize the true quasiparticle excitations and the role of
magnetic correlations. It was shown that the physics of two-impurity
Anderson model can be understood in terms of competition between of
itinerant motion of carriers and magnetic correlations of the RKKY
nature. The correct functional two-pole structure of the propagator has
been found for the strongly correlated case. This issue is still very
controversial~\cite{fye74} and the additional efforts must be
applied in this field.\\
The application of the IGF method to the theory of magnetic
semiconductors was very succsessfull~\cite{mar22}, \cite{mar23}.
As a remarkable results of our approach let me mention the finite
temperature generalization of the Shastry-Mattis theory for
magnetic polaron~\cite{mar23}, which clarified greatly the true nature
of the carrier in magnetic semiconductors. There are some analogy of
the Kondo-lattice type of model in ~\cite{mar22} with the
Kondo-Heisenberg model of copper oxides, however the physics are
different. There is a dense system of spins interacting with
smaller concentration of holes in HTSC. The application of IGF method to
spin-fermion model~\cite{kuz75} has been done by using the theory
of Heisenberg antiferromagnet~\cite{kuz26} and
allows one to consider carefully the
true nature of the carriers in $CuO_{2}$ planes.\\
And finally,  the new interesting application of the IGF method for
consideration of dynamics of quasiparticles and dynamical conductivity
of single electron resonant tunneling systems has been done recently
in papers~\cite{val76}, \cite{val77} (c.f.~\cite{mae78}). This reformulation
of IGF method has much in common with the approach of paper~\cite{yul79}.
\section{Conclusions}
In the present paper we have formulated the theory of the correlation
effects using the ideas of the quantum field theory for the interacting
electron system on a lattice. The main achievement of this formulation is
the derivation of the Dyson equation for two-time thermodynamic retarded
Green's Functions instead of causal ones. Such a formulation permit to
use the convenient analytical properties of retarded and advanced GF and
advantage of using the formal solution of the Dyson equation, which, in
spite required approximations for the self-energy, provide the correct
functional structure of the single-electron GF. This strong point of our
approach do not give the possibility of direct application of it to
the calculation of the two-particle GFs.
In this paper we have considered in details the idealized single-band
Hubbard model, which is one of the simplest (in the sense of formulation,
but not solution) and most popular model of correlated lattice fermions.
We have presented here the novel method of calculation of the quasiparticle
spectra for this model, as the most representative example. We hope that
this
explanation have been done with sufficient details to bring out their scope
and power, since we believe that such techniques will have application to
a variety of many-body systems with complicated spectrum and strong
interaction, as it was shown in Section 7. \\
In summary, with using IGF method we were able to obtain the closed self-
consistent set of equations determining the electron GF and self-energy.
These equations define the renormalization coefficient of the one-electron
GF\cite{mah7}, defined for a point (k, $\omega = \epsilon(k)$):
\begin{equation}
Z(k) = \frac{1}{(1 - {dM(k, \omega)\over d\omega})_{\omega = \epsilon(k)}}
\end{equation}
The renormalization coefficient (87) is an one of the most important
notion for the characterization of the single-particle behaviour of the
quasiparticle excitations in correlated many-body systems. For the Hubbard
model, these equations give the general microscopic description of
correlation effects for both the weak and strong Coulomb correlation,
determining of the complete interpolation solution of the Hubbard model.
Moreover, this approach gives the workable scheme for the definition
 of the relevant Generalized Mean Fields written in terms of appropriate
correlators. The most important conclusion to be drawn from of the present
consideration is that the GMF for the case of strong Coulomb interaction
have a quite non-trivial structure and cannot be reduced to the mean-density
functional. This last statement resemble very much the situation with the
strongly nonequilibrium system, where the single-particle distribution
function only not enough to describe the essence of the strongly
nonequilibrium state and more complicated correlation functions must be
taken into account, in accordance with general ideas of Bogolubov and
Mori-Zwanzig.
The IGF method is intimately related to the projection
method in this sense, which express the idea of a ``reduced description"
of the system in the most general form. This line of consideration are
very promising for developing the complete and self-contained theory
of the strongly interacting many-body systems.

\end{document}